%% file: Monier.tex
\documentclass[12pt,preprint]{aastex}

\newcommand{\etal}{\mbox {\it et al.}}

\newcommand{\lya}{\mbox{\,Ly$\alpha$}}

\newcommand{\Mftf}{\mbox{\,M$_{1450}$}}

\begin{document}

\title{The BTC40 Survey for Quasars at $4.8 < z < 6$\altaffilmark{1}}

\altaffiltext{1}{Based on observations performed at the Blanco 4-m telescope at the Cerro Tololo 
Inter-American Observatory and the Anglo-Australian Telescope}

\author{Eric M. Monier\altaffilmark{2,3,4},
Julia D. Kennefick\altaffilmark{4,5}, Patrick B. Hall\altaffilmark{4,6}, Patrick S. Osmer\altaffilmark{3}, \\
Malcolm G. Smith\altaffilmark{7}, Gavin B. Dalton\altaffilmark{5}, 
\& Richard F. Green\altaffilmark{4,8}}

\altaffiltext{2}{monier@astronomy.ohio-state.edu}

\altaffiltext{3}{Department of Astronomy, The Ohio State University,
140 W. 18th Ave., Columbus, OH  43210}

\altaffiltext{4}{Visiting Astronomer, CTIO, National Optical Astronomy 
Observatories, which is operated by the Association of Universities for 
Research in Astronomy, Inc.  (AURA) under cooperative
agreement with the National Science Foundation.}

\altaffiltext{5}{NAPL, Keble Road, Oxford OX1 3RH, England, UK}

\altaffiltext{6}{Princeton University Observatory, Princeton, NJ 08544-1001, and Pontificia
Universidad Catolica de Chile, Departamento de Astronom\'{i}a y
Astrof\'{i}sica, Facultad de F\'{i}sica, Casilla 306, Santiago 22, Chile}

\altaffiltext{7}{Cerro Tololo Inter-American Observatory, Casilla 603, La Serena, Chile}

\altaffiltext{8}{National Optical Astronomy Observatories, P.O. Box 26732, Tucson, AZ 857726}

\slugcomment{Accepted for publication in the {\it Astronomical Journal}}

\begin{abstract}

The BTC40 Survey for high-redshift quasars is a multicolor search using
images obtained with the Big Throughput Camera (BTC) on the 
CTIO 4-m telescope in $V$, $I$, and $z$ filters to search for
quasars at redshifts of $4.8 < z < 6$.  The survey covers 40 deg$^2$ in $B$, $V$, \&
$I$ and 36 deg$^2$ in $z$.  Limiting magnitudes (3$\sigma$) reach
to $V = 24.6$, $I = 22.9$ and $z = 22.9$.  We used the ($V-I$) vs. ($I-z$) two-color
diagram to select high-redshift quasar candidates from the objects classified
as point sources in the imaging data.  Follow-up spectroscopy 
with the AAT and CTIO 4-m telescopes of candidates having $I \leq 21.5$ 
has yielded two quasars with redshifts of 
$z = 4.6$ and $z = 4.8$ as well as four emission line galaxies with $z \approx 0.6$.  
Fainter candidates have been identified down to $I = 22$ for future spectroscopy on 
8-m class telescopes.

\vskip 1.5in

\end{abstract}


\section{Introduction}

Surveys for faint quasars at $z > 4.5$ and the subsequent constraints they
place on the quasar luminosity function (QLF) will eventually determine how
luminosity evolution and density evolution each contribute to the declining
space density of quasars established for $3 \lesssim z \lesssim 4.3$ (Warren, Hewett, \& Osmer 1994, 
hereafter WHO; Schmidt, Schneider, \& Gunn 1995, hereafter SSG; Kennefick, Djorgovski, \& DeCarvalho 1995)
and now observed out to redshifts of $z\sim5$ (Fan \etal\ 2001b).  The QLF in turn
is an important input to models of structure formation in the early Universe
(Haehnelt, Natarajan \& Rees 1998).

Additionally, high-redshift quasars provide insight into the nature of quasars and
their environments in the early Universe (Haehnelt \& Kauffmann 2000),
contribute to the ionizing UV background (Madau, Haardt \& Rees 1999), 
and act as background illumination for absorption-line studies of the 
intergalactic medium.  It is therefore important to continue to search
for quasars of all luminosities at the highest possible redshifts, and thus the
earliest possible epochs.

We began this survey to address the shape of the QLF at redshifts $z \gtrsim 5$.
WHO found evidence that
the positive evolution in the QLF at $0 < z < 2.2$ continues to $z\approx 3.3$
and then the space density declines by a factor of 6.5 at $z = 4$.  SSG
found that space densities have a maximum
between $z=1.7$ and $2.7$ and then decrease by a factor of 2.7 per unit
redshift beyond $z=2.7$.  Extrapolations of these QLF's to $5 < z < 6$ predict
0.02 (WHO) to 0.6 (SSG) quasars per deg$^2$ to $I=22$.  

The major effort to find quasars at higher redshifts is the 
Sloan Digital Sky Survey (SDSS), which continues to be remarkably successful 
at finding $z > 4$ quasars.  SDSS has discovered more than 100 such objects
including one at $z = 6.3$, the most distant published (see, e.g. Fan
\etal\ 2000, 2001c; Anderson \etal\ 2001).  However, SDSS is limited to 
$z > 4.5$ quasars with M$_B \lesssim -26$ and misses the bulk of the
population, which is found at lower luminosities.  Thus, there is a need
for surveys to find less luminous $z > 4.5$ quasars to address the shape of the faint
end of the QLF.  Sharp \etal\ (2001) have 
presented initial results from one such survey, finding two $z > 4.5$ quasars
to $i \approx 21.5$ in 10 deg$^2$ of $griz$ data.  

The BTC40 survey is a deep, 40 deg$^2$ survey in $BVIz$ filters undertaken to search
for clusters of galaxies, morphologically-selected gravitational lenses,
and quasars at $z \gtrsim 4.8$.  The results on clusters and gravitational lenses 
will be presented elsewhere.  In this paper we present results of our efforts 
using a 4-m telescope and large format camera to complement the SDSS quasar 
search and extend it to lower luminosities.  

We used the $VIz$ imaging data to select quasar candidates over
36 deg$^2$ of sky down to $I \leq 22$, corresponding to absolute magnitudes of $M_B \lesssim -24.7$
at $z=4.8$.  The selection process compared the expected colors of quasars at redshifts $4.8 < z < 6$
to the locations of catalogued stellar objects in ($V-I$) vs. ($I-z$) color space.
Follow-up spectroscopy at the CTIO 4-m and AAT was attempted to $I < 21.5$ and has resulted in the 
discovery of two quasars with redshifts of $z = 4.6$ and $z = 4.8$, as well as several emission-line 
galaxies at $z \approx 0.6$.  Spectroscopy of the fainter candidates, down to $I=22$, will be the 
focus of our future efforts on larger telescopes.

We describe the survey imaging data in $\S$2, the candidate selection in $\S$3, 
and the follow-up spectroscopy in $\S$4.  We discuss our results in $\S$5.

\section{Imaging}

\subsection{Observations}

The Big Throughput Camera (Tyson \etal\ 1992; Wittman \etal\ 1998) contains a
2$\times$2 array of thinned 2048$\times$2048 SITe CCDs with 24$\mu$m pixels.  
Used at the prime focus of the CTIO 4-m Blanco Telescope (and since replaced
by the MOSAIC camera), the BTC plate scale is 0\farcs43 pixel$^{-1}$,
resulting in a sky area of 14$\farcm$7 $\times$ 14$\farcm$7 imaged by each CCD.
The CCDs are separated by 5$\farcm$4, for non-contiguous
coverage of 0.24 deg$^2$ per pointing.

Six survey fields (Table \ref{fields}) were selected with declinations to provide low
airmass, and $|b|$ chosen to minimize Galactic \ion{H}{1} and contamination by stars.
The survey was split between spring and fall data sets, each containing three
fields separated by $\sim$2.5 hours in right ascension.

Initial BTC imaging in the Johnson-Cousins $BV$ and Kron-Cousins $I$ filters was performed
at the Blanco 4-m telescope at CTIO in the spring and fall semesters of 1997 (Table \ref{journalofimaging}).   
A ``lawn-mowing'' pattern was used, in which the telescope moved back and forth first 
in right ascension and then in declination.  An overlap of 1$\arcmin$ in both directions provided 
continuity between adjacent pointings for later ``bootstrapping'' of the photometry.  

The $V$ and $I$ data were reduced and catalogs produced (both procedures are described below
for the full data set) and $\approx$200 objects with ($V-I$) $>$ 4.0 were selected as candidates
for $z > 5$ quasars.  The $B$ data were not used for the quasar survey 
because the $B-V$ color provides no useful information for finding high-redshift quasars; 
the 912 \AA\ Lyman limit is well into the $B$ filter by $z = 4.8$, while the flux in $V$ is 
heavily depressed by the \lya\ forest.
Spectroscopy was performed at the CTIO 4m on 20 of the brighter candidates 
under poor conditions in 1998 April.  All of these candidates turned out to be late-type
(M5 or later) stars.  In light of this result, we resolved to improve the survey efficiency
and reduce cool-star contamination of the quasar candidate sample by obtaining follow-up
imaging in the near-infrared $z$ filter.

\begin{figure}
\centerline{\includegraphics[width=5in]{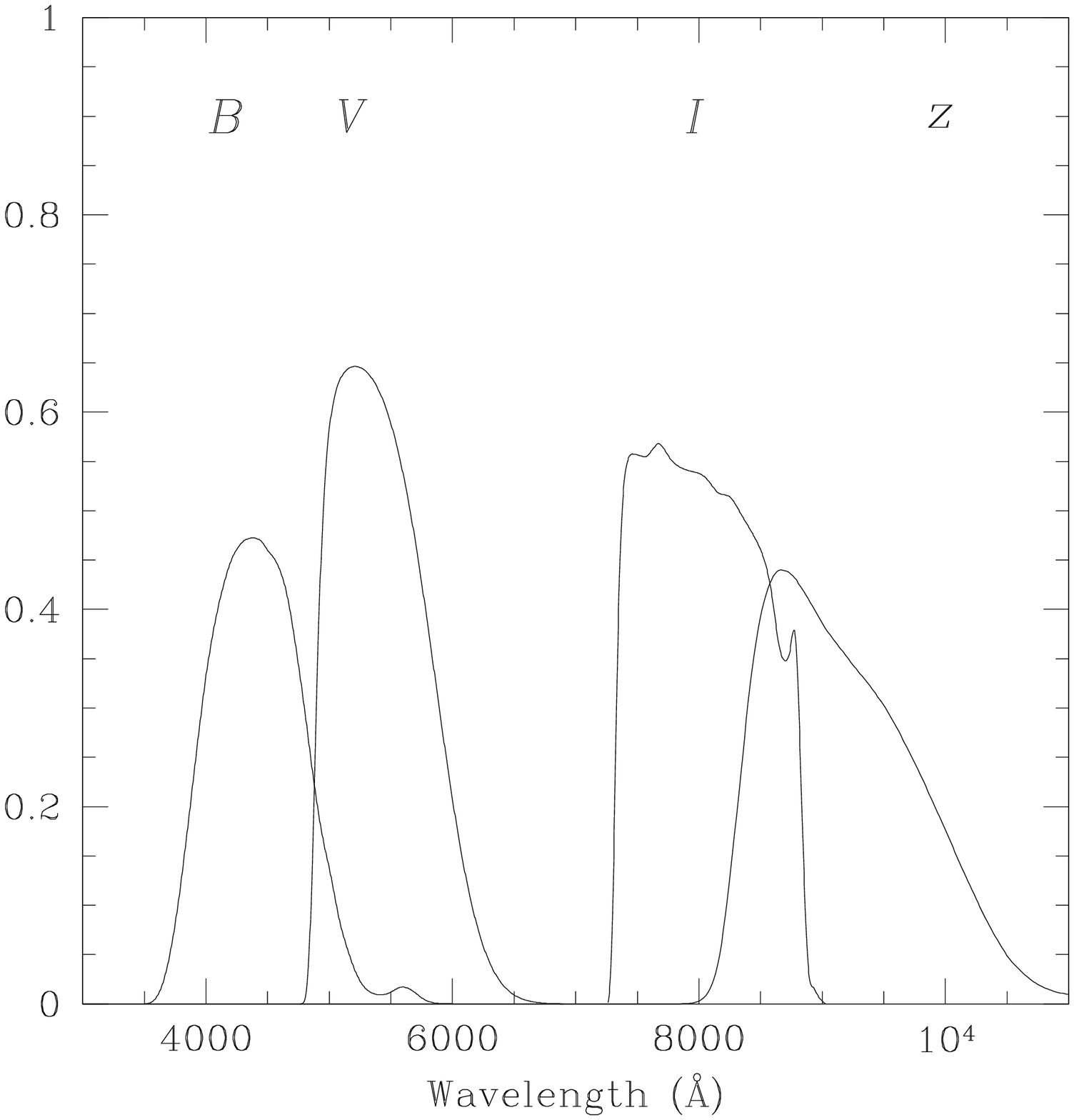}}
\figcaption[Monier.fig1.ps]{Response curves of the four filters
used in the BTC40 survey, incorporating the filter response
and CCD response.\label{filt}}
\epsscale{0.65}
\end{figure}

The $z$ filter in use at CTIO is a longpass filter matching the response of the $z^\prime$ filter in the SDSS 
(Fukugita \etal\ 1996; Gunn \etal\ 2001).  The filter uses RG830 Schott color glass to provide the
short-wavelength cutoff, and the long-wavelength response is determined entirely by the decreasing 
sensitivity of the CCD past 9000 \AA.  The combined CCD and filter response curves are shown
in Figure \ref{filt}.  By extending coverage into the near-IR past $I$, the $z$ filter 
provides a means of distinguishing quasars from late-type stars.
At redshifts of $z \approx 5$, the quasar \lya\ $\lambda$1216 emission line moves into 
the $I$ filter at $\approx$7250\AA.  Quasars thus become progressively bluer in ($I-z$) 
than late-type stars of similar ($V-I$) 
(see \S3), allowing for the separation of the quasars from the stars in color-color space.
Imaging of the survey fields in the $z$-band was performed in 1998 November 
and 1999 February (Table \ref{journalofimaging}).  \footnote{CCDs 1 and 2 of the BTC 
were replaced in 1998 May, between the $BVI$ and the $z$ runs.}

\subsection{Data Reduction}

We processed the imaging data by using the CCDRED task in the IRAF\footnote{IRAF is distributed by the National 
Optical Astronomy Observatories, which are operated by the Association of Universities 
for Research in Astronomy, Inc., under cooperative agreement with the National
Science Foundation.} software package to correct the object and calibration frames from each observing run 
for overscan and bias.
For each CCD chip, we median-combined and normalized the dome flats from each night and
used them to remove pixel-to-pixel variations in the field and standard-star frames.  
We then took all of the field images from the run and
median-combined them to create ``super sky'' flats for the four CCD chips.  After smoothing these flats we
used them as illumination corrections to remove the large-scale variations across each chip.  
Finally, to correct for fringing in the $I$ and $z$ data, we took the unsmoothed super-sky flats,
scaled them with the {\it defringe}\footnote{Part of Pat Hall's Add-on Tasks (PHAT) under IRAF.} 
task to the level of the individual object frames, and subtracted them from those frames.

\subsection{Catalogs}

We processed the reduced images with the SKICAT software 
package (Weir \etal\ 1995) to generate our object catalogs.  Briefly, SKICAT uses FOCAS 
(Jarvis \& Tyson 1981; Valdes 1982) for object detection, photometry, and 
object classification.
The resulting catalogs reside in an online SYBASE database 
and are easily accessed through the SKICAT interface.  Queries can be performed on the 
database to, e.g., select point sources within a given magnitude range.  Our database 
contains approximately 476,000 stellar objects with $I \geq 16$ down to the 3$\sigma$ 
limits ($V=24.6$, $I=22.9$, $z=22.9$) of the survey.

An important component of SKICAT is the ability to match features between catalogs 
based on their measured positions.  For this survey, our objective was to construct 
matched $VIz$ catalogs of objects classified as point sources in the $I$ filter and use them to create 
($V-I$) vs. ($I-z$) two-color diagrams of the survey area, as described in \S3.
Construction of the matched catalogs relies on the object coordinates, therefore the astrometry
of the objects must be accurate.

\subsection{Astrometry}

The large field of view of the BTC results in  images with significant geometric distortions 
due to sky curvature, slight rotations between the chips and distortions of the prime focus 
corrector (Wittman \etal\ 1998).  
Therefore, accurate astrometry of objects catalogued in this survey had to be 
established before matching could be performed between filters.  We used the {\it undist} 
ray tracing program (written by I. Dell'Antonio) in a two-step process to correct the positions 
contained in the object catalogs.  For a given pointing (including RA, Dec, hour angle and filter) 
the program generates a grid of CCD $x$,$y$ pairs
with corresponding RA and Dec, based on the known optical properties of the BTC and several input
parameters (field center, hour angle, filter, and date of observing run).  The IRAF tasks 
{\it geomap} and {\it geoxytran} use the output spatial information 
to compute and perform the transformation from the FOCAS $x$,$y$ coordinates of cataloged
objects into RA and Dec values.

A first run through this process produced an initial correction to the object coordinates.
We then downloaded USNO catalogs (Monet \etal\ 1998) of the stars in the areas of the sky 
covered by each pointing and compared the coordinates of stars common to the USNO and survey 
catalogs to produce a set of offsets 
ranging from a few arcseconds up to $\sim$2\arcmin\ for some fields.
We used these  as the input to a second pass of the IRAF routines to refine the transformation, and from this
result calculated new object coordinates and updated the database.  The RMS residuals on the fit
to the USNO coordinates were on the order of $\approx$0.4 arcseconds.

\subsection{Photometry}

Conditions during the imaging runs were generally not photometric.   To calibrate the 
data we determined initial rough zero points for the data from standard stars 
observed at the beginning and end of each night.  In an initial test
of the candidate selection criteria (\S3), the position of the stellar locus in 
color-color space fluctuated from field to field, indicating changing conditions 
over the course of the run and the individual nights.
Portions of some nights may have been photometric, but
standard-star observations obtained under photometric conditions were not 
extensive enough to allow a precise photometric solution for the full data set.  
However, note that when selecting high-redshift quasar candidates 
only accurate differential photometry is crucial.  Candidates are chosen by where they lie with respect to
the stellar locus in color-color space, making absolute photometry less critical.  

Nevertheless, we took several steps to ensure the accuracy of the
absolute photometry.  We created a model stellar locus using the Bruzual-Persson-Gunn-Stryker 
(BPGS) spectrophotometric atlas available in the STSDAS/SYNPHOT package under IRAF.
This atlas is an extension of the Gunn-Stryker optical atlas (Gunn \& Stryker 1983) into the UV
and also includes infrared data from Strecker, Erickson, \& Whittenborn (1979).
The stellar locus results from folding the colors through the combined CCD/filter response 
and plotting the resulting ($V-I$) vs. ($I-z$) colors.   After creating
and examining color-color plots of ($V-I$) vs. ($I-z$) for individual pointings,
we selected a pointing from the spring dataset as having been taken under 
near-optimal conditions based on the tightness of the observed stellar locus and its overlap
of the model stellar locus.  
We determined the shifts in color-color space needed to put each pointing onto the same scale 
by using a light table and evaluating the necessary shifts by eye.  We eliminated four pointings 
of Field 6 from candidate selection based on the low numbers of objects 
and the degree of scatter in the resulting stellar locus.  Visual inspection of these 
fields and the observing logs indicate conditions were especially poor at the time of these 
observations.

\begin{figure}
\centerline{\includegraphics[width=5in]{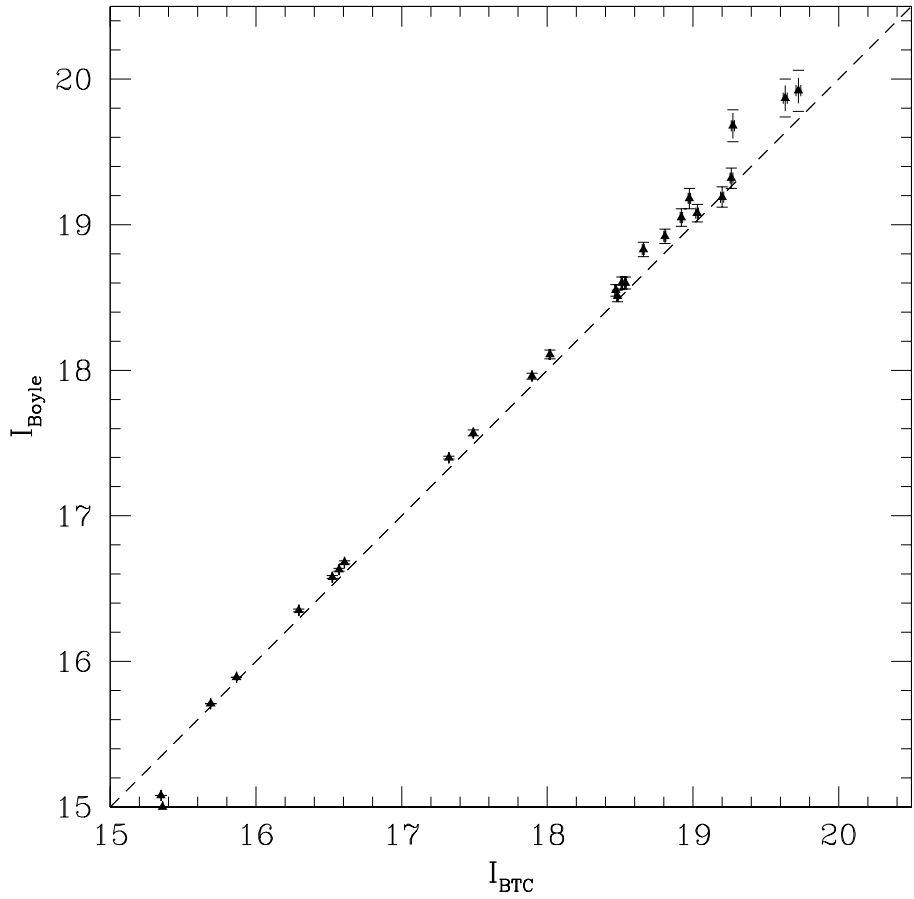}}
\figcaption[Monier.fig2.ps]{Comparison of the $I$ magnitudes of 
stellar objects common to the BTC40 survey and Field 866 of 
Boyle \etal\ (1995).  The two brightest objects in the lower left-hand
corner are saturated in the BTC40 data.\label{boyle}}
\epsscale{1.0}
\end{figure}

One of the survey fields was originally chosen to overlap part of the faint photometric
calibration sequence of Boyle \etal\ (1995).  We compared our final photometry of the sources
in this 6 $\times$ 6 arcminute overlap region to the values from the catalogs of Field 866 in Boyle \etal. 
The result is shown in Figure \ref{boyle}.  The magnitudes agree well for $I < 19$, neglecting the
two brightest objects, which are saturated in the BTC40 data.  At the fainter end, however,
the figure suggests a discrepancy of several tenths of a magnitude between the two 
measurements.  We intend future observations of the survey fields to refine the survey photometry, but,
as noted above, the relative photometry was adequate for selecting high-redshift quasar candidates.

The journal of observations in Table \ref{journalofimaging} includes the average 3- and 5-$\sigma$ magnitude limits 
reached in each survey field.  The average 
limiting magnitudes (5$\sigma$) for the imaging data over the entire survey 
are $V=24.0$, $I=22.4$, and $z=22.4$ (AB$_{95}$).  Note that the original BTC CCD chip \#1 was less 
sensitive than the other three, especially in the blue (Wittman et al 1998).  
As a result, the 5$\sigma$ limiting 
magnitudes for chip \#1 in $V$ are ${\Delta}m_{lim,{\rm V}}\approx$ 0.1 - 0.3 brighter 
than those of chips \#2, \#3, \& \#4.  The chip \#1 sensitivity in the $I$ filter was less 
discrepant, keeping any effect on the limiting magnitude ${\Delta}m_{lim,{\rm I}}\leq$0.1 for
chip \#1 $I$ data.  Because the $V$ data are much deeper than the $I$ data, and 
candidate selection was limited to $I \leq 21.5$ variations in chip sensitivity ultimately did
not influence candidate selection.

\section{Candidate Selection}

\begin{figure}
\epsscale{1.0}
\centerline{\includegraphics[width=7.0in]{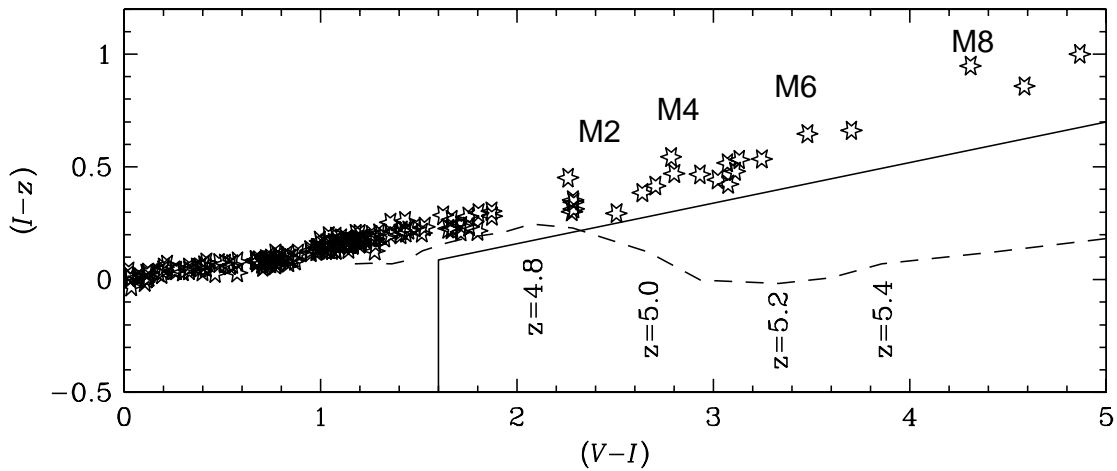}}
\vskip -5.0in
\figcaption[Monier.fig3.ps]{Plot of ($V-I$) vs. ($I-z$) colors showing the expected
location of high-redshift quasars in color-color space, below and to the
right of the solid selection lines.  The dashed track follows
the expected colors of quasars at redshifts from 4 to 6, with the colors 
of several redshifts marked.  These colors were computed from 10 synthetic
quasar spectra of varying continuum slopes and realizations of the \lya\
forest in 0.1 redshift bins.  At quasar redshifts of $z \sim 5$, quasars become
progressively bluer in ($I-z$) than the late-type stars of similar ($V-I$),
which enables the separation of the quasars from the stars in color-color space.
The model stellar locus was calculated from the Bruzual-Persson-Gunn-Stryker
spectrophotometric stellar database.\label{models}}
\end{figure}

The general region of ($V-I$) vs. ($I-z$) color space in which $4 < z < 6$
quasars will be found is readily seen in Figure \ref{models}, a plot of the 
stellar sequence taken from the BPGS  stellar atlas in the STSDAS/SYNPHOT package. 
The quasar colors are from synthetic quasar spectra as described in Kennefick \etal\ (1996).  
We used this predicted region as a guide to selecting initial candidates from ($V-I$) vs. ($I-z$) diagrams
of BTC40 objects classified as point sources (FOCAS 'stars' or 'fuzzy stars') in the $I$-band data.
We performed a visual inspection of the images of those candidates to ensure the colors 
had not been skewed by cosmic ray hits or cosmetic flaws in the CCDs.  Finally, a priority was
assigned to each candidate based on the magnitude, location in color-color space, and
the visual inspection.

The selection process for our initial spectroscopy runs used magnitudes as calculated in
SKICAT using FOCAS as described by Weir \etal\ (1995).  FOCAS calculates two magnitudes: an 
aperture magnitude calculated within a specified radius (we used three pixels during SKICAT processing), 
and a `total' magnitude based on an area of twice the detection area.
As discussed by Weir \etal\ (1995), the total magnitude
carries the associated risk of increased random error.  After our first observing run in 1999 November at
CTIO, in which only one observed candidate turned out to be a quasar, we compared the FOCAS photometry 
to magnitudes measured independently with PHOT in IRAF.  We concluded that for faint point sources
the random errors in the SKICAT/FOCAS measurement can be significant, moving stars out of the stellar
locus and into the region of quasar candidate selection (and presumably vice versa).  

\begin{figure}
\epsscale{1.0}
\centerline{\includegraphics[width=6.5in]{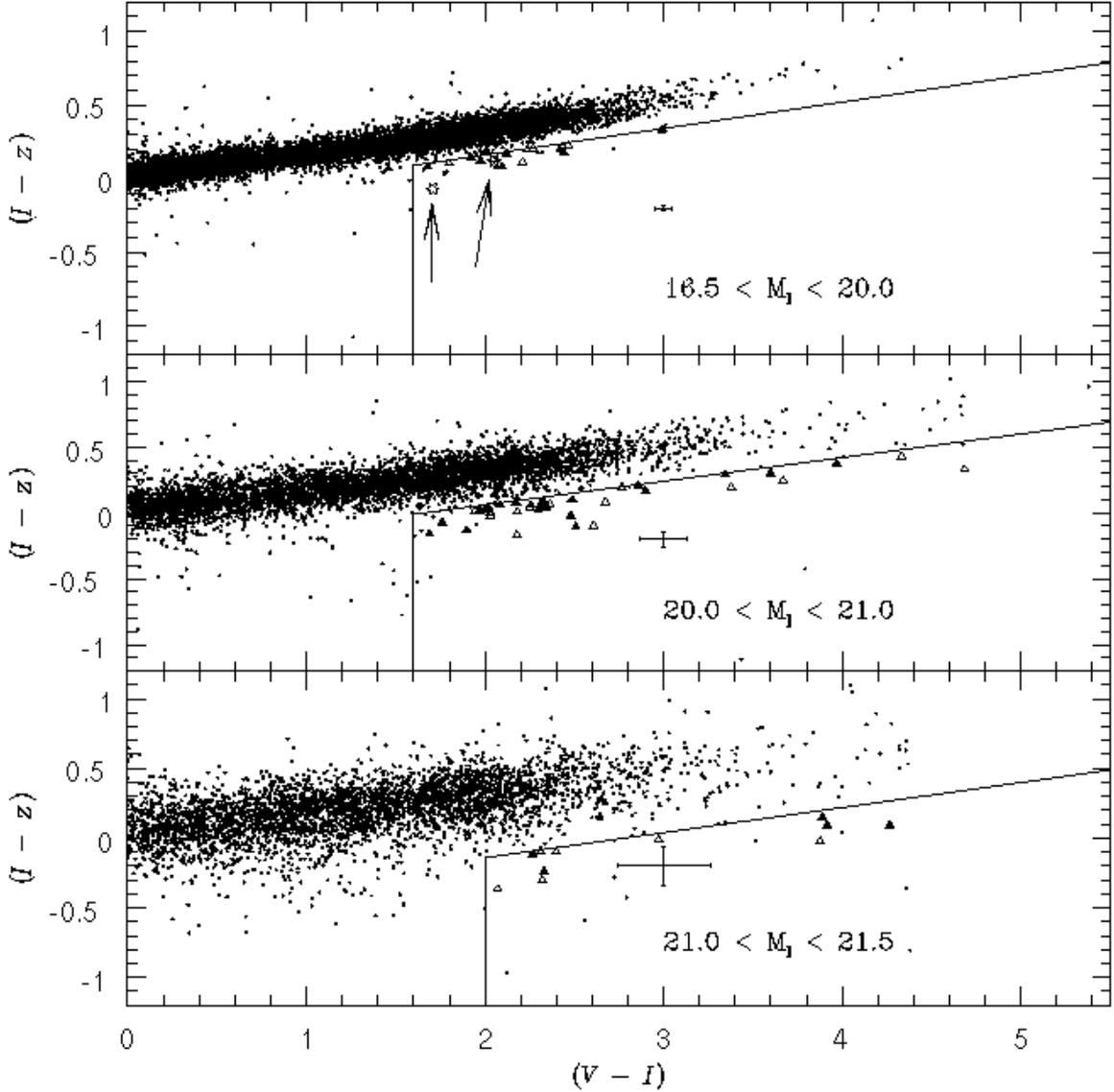}}
\figcaption[Monier.fig4.ps]{($V-I$) vs ($I-z$) plots of stellar objects in 4.75 deg$^2$ (Field 3) 
of the survey overplotted with triangles to indicate the objects selected as
high-redshift quasar candidates.  Filled triangles represent
candidates we observed spectroscopically.  The two quasars discovered are the open
stars indicated by arrows in the top panel.  The CNELGs are the three filled triangles 
with ($V-I$)$\approx$1.7 in the top and middle panels.   Typical errors on the photometry are plotted 
for ($I-z$)$ = -0.2$ and ($V-I$)$ = 3.0$.  Dots appearing in the selection areas are 
due to bad CCD columns or cosmic rays and were eliminated by visual inspection of the $VIz$ 
images during the candidate selection process.\label{colors}}
\end{figure}

Therefore, we re-measured the magnitudes using the IRAF PHOT package and an aperture radius of two 
pixels (roughly twice the seeing) and proceeded to compile candidate lists based on the colors
resulting from the IRAF photometry.  
Objects that were also in the selection area using FOCAS colors were given higher priority for
follow-up spectroscopy.  The initial spectroscopy runs allowed us to refine the selection criteria 
and estimate how close to the stellar locus we could 
reasonably hope to go.  This determination requires balancing the number of candidates selected with
the increasing likelihood of contamination by the cool stellar population as one moves closer to the stellar locus.  
Figure \ref{colors} shows the stellar locus from a portion of the survey data overplotted with the complete 
candidate set broken down by magnitude range.  The objects in the region below and to the right of the lines satisfy the
criteria used in selecting candidates.  

\section{Spectroscopy}

We performed follow-up spectroscopy of selected candidates 
during three observing runs at the CTIO 4m and three at the AAT (Table \ref{journalofspec}).
Over the course of these six runs we were able to observe 82 quasar candidates.
However, many of these are not included in our later survey sample of sixty-two candidates with 
$I{\leq}21.5$, due to (1) the adoption of IRAF photometry after the first observing run and (2) 
modifications to the selection criteria based on results of the initial runs.
We observed 40 objects from the list of 62 candidates, 27 at CTIO and 13 at the AAT; 
several promising candidates with inconclusive initial spectra were observed at both telescopes.
In this section we describe the setup at the two locations and the data reduction process.

\subsection{CTIO}

We used the RC Spectrograph at the CTIO 4m with similar setups on 1999 November 14-16, 2000 May 9-11, 
and 2000 October 16-17.  The RC Spectrograph on the CTIO 4m uses the Blue Air Schmidt (BAS) camera 
and a Loral 3K CCD, which we formatted to 3071$\times$800 for faster readout.  
We chose the G181 grating (316 lines/mm) and the GG495 blocking filter 
for a resolution of 2\AA\ pixel$^{-1}$ over the range $\approx$5000 - 11000 \AA.
During good seeing conditions we maintained the slit width at 1$\arcsec$ and opened it
to 1$\farcs$5 when the seeing deteriorated.  Exposure times ranged from 900s to 3600s, depending 
on the brightness of the candidate.  We observed 27 candidates from our sample over the 
course of the three runs, down to a cutoff magnitude of $I \simeq 21.5$. 

\subsection{AAT}

We performed the spectroscopy at the AAT using LDSS++ with the 165 \AA/mm grating
for a dispersion of 2.6\AA\ pixel, providing coverage over $\sim$5200-10000\AA.  
We used a longslit 1\arcsec\ wide when possible,
but in general we were limited by unfavorable conditions to a slit width of 1\farcs7.  LDSS++ can 
be operated in ``nod and shuffle'' mode (Glazebrook \& Bland-Hawthorn 2001), in which
the telescope is nodded rapidly by 10--20\arcsec\ while the spectra are 
recorded on two adjacent regions of the MIT Lincoln Lab (MITLL) CCD 
through charge shuffling.  The method allows optimal sky subtraction and is ideal for extracting
fainter targets.  Conditions were fair to poor for most of three observing runs, and we 
observed 13 candidates from our survey list.

\subsection{Data Reduction}

We reduced the candidate spectra with CCDRED under IRAF, including subtracting the overscan region and 
flatfielding the data using quartz flats obtained during the runs.  For the nod-and-shuffle 
AAT data we performed the additional step of subtracting the two exposed regions of the CCD, 
effectively eliminating the night sky emission lines.  We used the IRAF `apall' task to select and size 
the apertures interactively and trace the spectra across the chip.  We also specified the regions used 
for background subtraction in the CTIO data.  We extracted arc calibration spectra by using traces 
obtained from standard-star spectra, used the 'identify' task on the arc spectra to obtain a dispersion solution,
and proceeded to apply this solution to the extracted spectra of the candidates.  Finally, we combined 
spectra of objects having multiple exposures.  We also used extracted standard-star spectra to 
get the shape of the object spectra and a rough flux calibration.

\subsection{Spectroscopic Results}

The positions, $I$ magnitudes, colors, and, where possible, identifications of 
the observed candidates are given in Table \ref{results}.
The list includes two quasars, ten stars, three compact narrow emission-line galaxies (CNELGs),
15 objects that could not be identified due to the low S/N of the spectra, and
9 objects with spectra too faint to be extracted (but possessing no obvious emission lines).

\subsubsection{Quasars}

\begin{figure}
\epsscale{1.0}
\centerline{ \includegraphics[angle=-90]{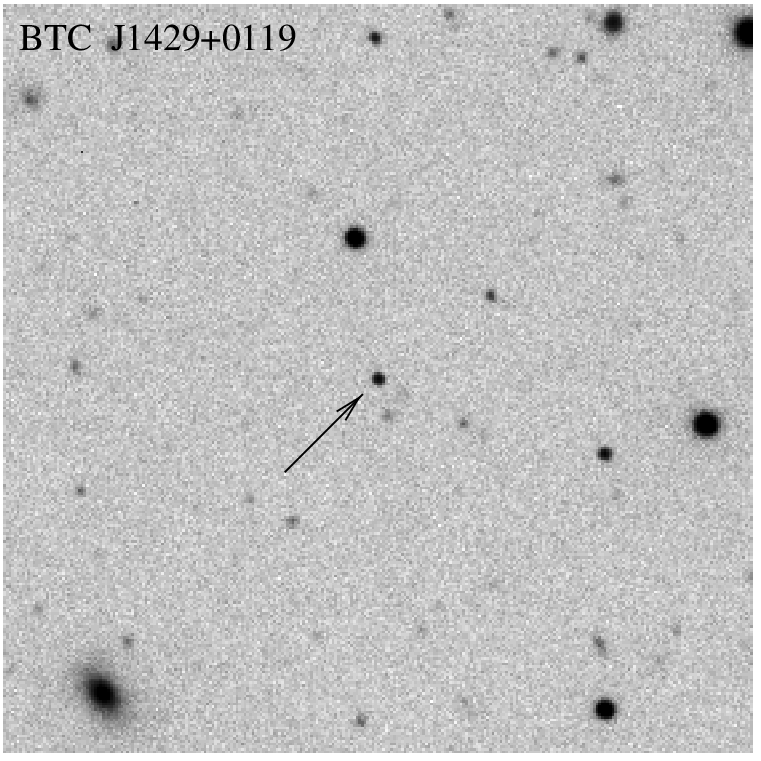}
\includegraphics[angle=-90]{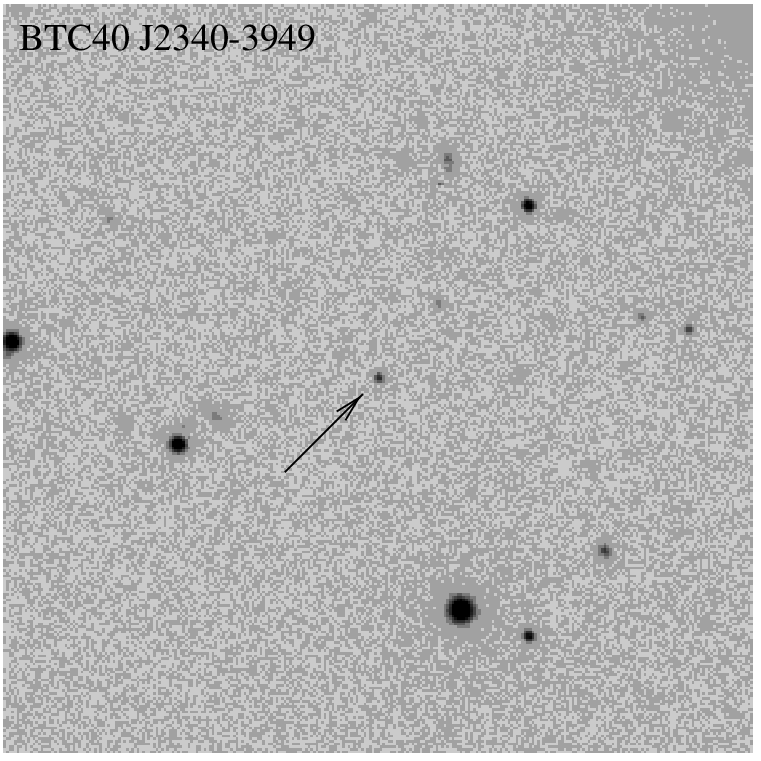}}
\figcaption[]{Finding charts for the two BTC40 quasars.  The quasars are marked by arrows
at the center, and the fields are 2\arcmin\ square.  North is up and east is to the left.
\label{qsochart}}
\end{figure}

\begin{figure}
\epsscale{1.0}
\centerline{\includegraphics[width=6in]{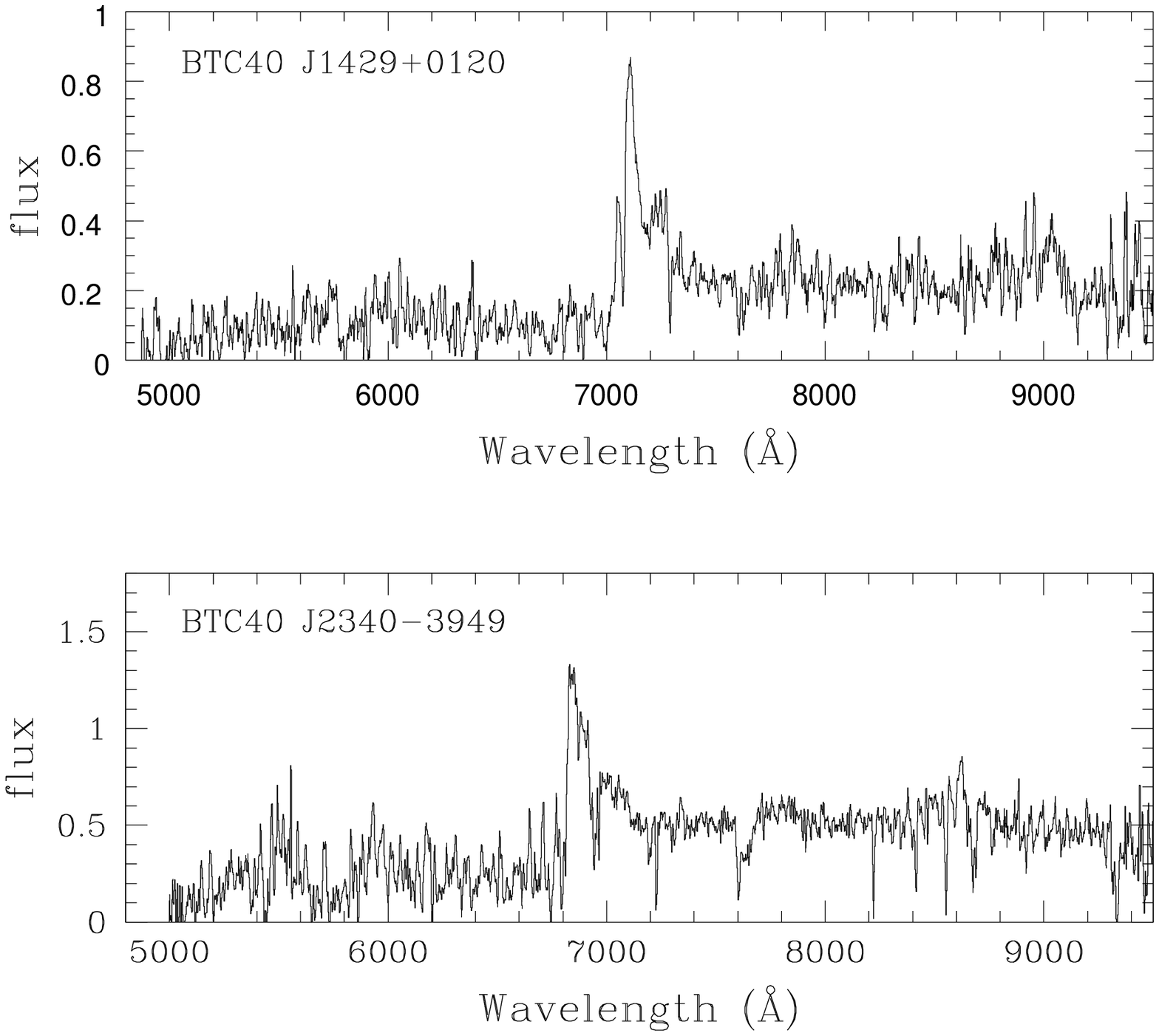}}
\figcaption[Monier.fig6.ps]{Spectra of the two quasars found in this survey, showing \lya\
emission at $\sim$7000\AA\ and continuum depression shortward of \lya\ due to the \lya\
forest.  BTC40 J2340$-$3949 may have weak BAL troughs (or strong associated absorption)
in \ion{N}{5} at $\approx6945$ \AA\ and \ion{C}{4} at $\approx8680$ \AA.  The absorption
at 7600 \AA\ is atmospheric.\label{qsos}}
\end{figure}

Two of the identified objects are quasars at redshifts $z\simeq4.6$ and $z\simeq4.8$ and
magnitudes of $I=19.4$, toward the brighter limit of our survey.  Finding charts for these
quasars are given in Figure \ref{qsochart}, and the spectra are presented in 
Figure \ref{qsos}.  Both quasars show strong \lya/\ion{N}{5}
emission and weaker emission due to \ion{C}{4}.  Continuum flux shortward of \lya\ is
depressed by the intervening \lya\ forest.

One of the quasars, BTC40 J2340$-$3949, was detected in the deep, 1.4GHz
ATESP radio survey (Prandoni \etal\ 2000) with a peak flux of 0.57 mJy.
We can adapt the radio-optical flux ratio, $R_{r-o}$, of Kellerman \etal\ (1989) for a 
high-redshift source, substituting the $I$-band (8000 \AA) magnitude for the 
$B$-band (4400 \AA) and scaling the radio flux to the proportionally lower frequency
$\nu$(8000/4400)$^{-1}$ by assuming a power law of the form $f\sim\nu^{-0.5}$.
In this case, $R_{r-o}\approx10$, putting the quasar just on the border of being 
classified as radio-loud.  BTC40 J2340$-$3949 may have a weak BAL trough (or 
strong associated absorption) seen in \ion{N}{5} at $\approx6945$ \AA\ and 
\ion{C}{4} at $\approx8680$ \AA\ (the absorption at 7600 \AA\ is atmospheric), and 
may therefore be similar to the radio-moderate BAL quasars discovered in large numbers 
by the FIRST Bright Quasar Survey (Becker et al. 2000).  

\subsubsection{Compact Narrow Emission Line Galaxies}

\begin{figure}
\centering {
\includegraphics[width=2in,angle=-90]{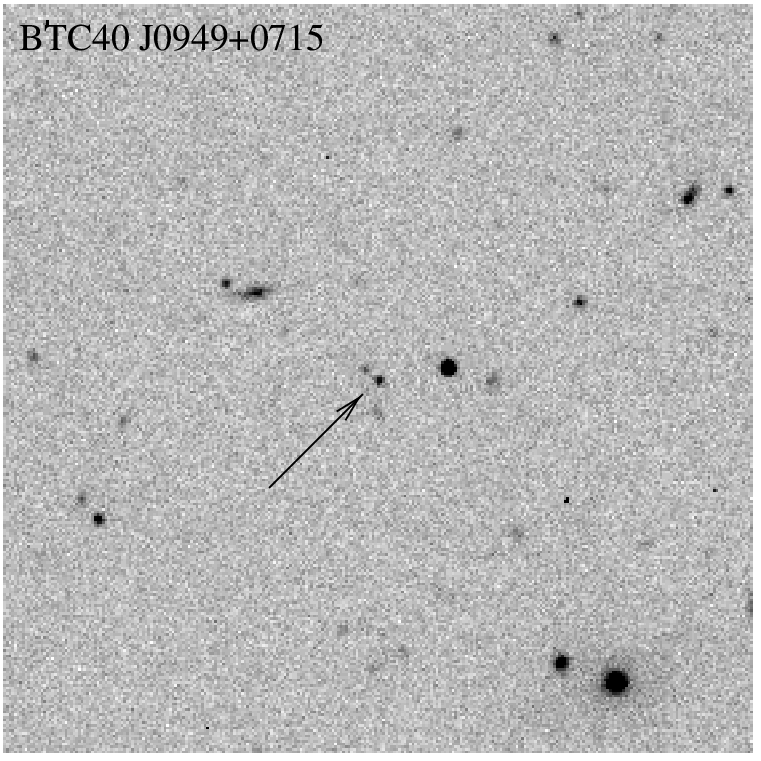} 
\includegraphics[width=2in,angle=-90]{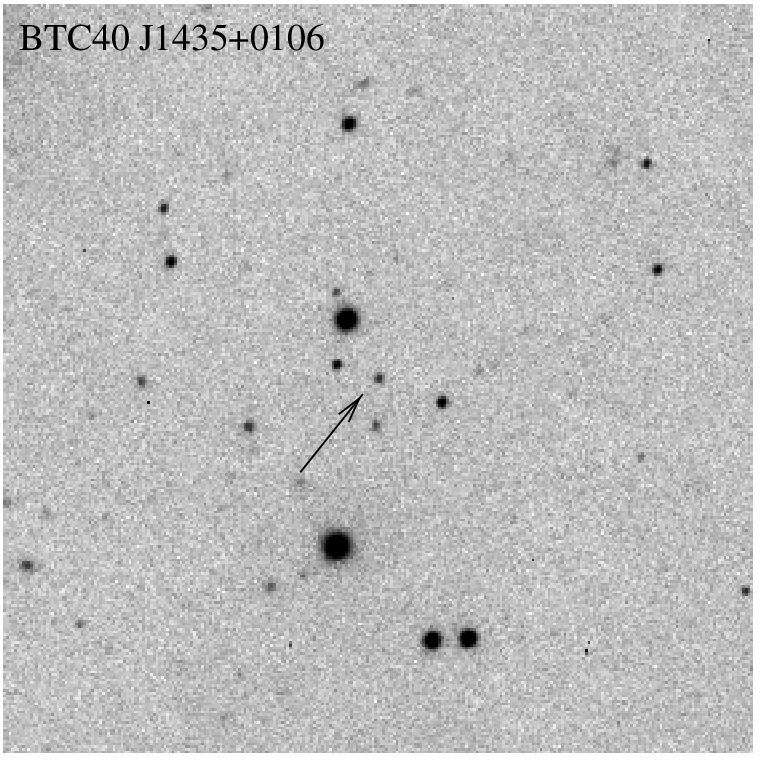} 
\includegraphics[width=2in,angle=-90]{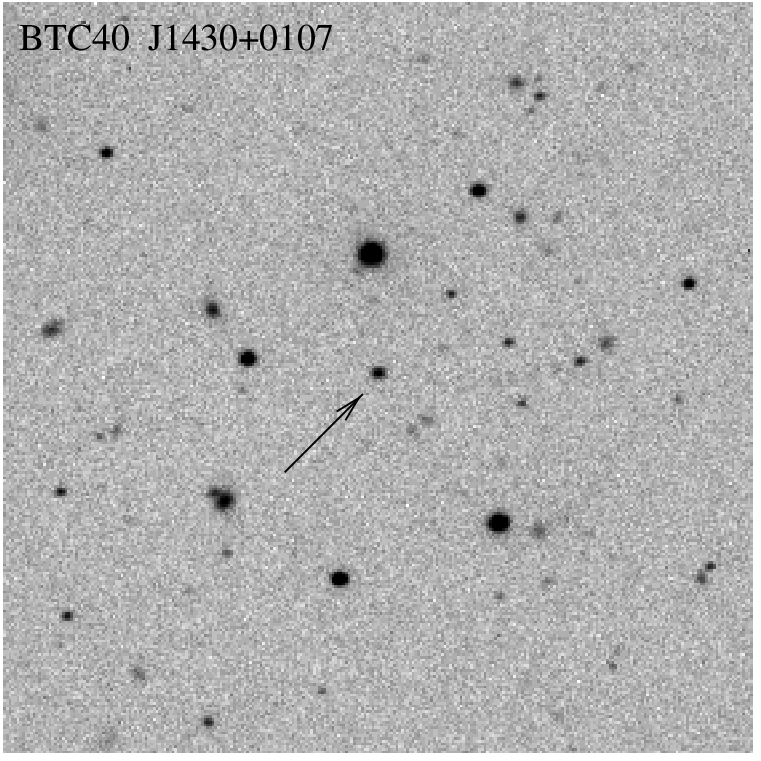}}
\figcaption[]{Finding charts for the three compact narrow emission line galaxies, marked
by arrows in the center.  The fields are 2\arcmin\ square.  North is up and east is to the left.
\label{cnelchart}}
\end{figure}

\begin{figure}
\epsscale{1.0}
\vskip -1.0in
\centerline{\includegraphics[width=6.5in]{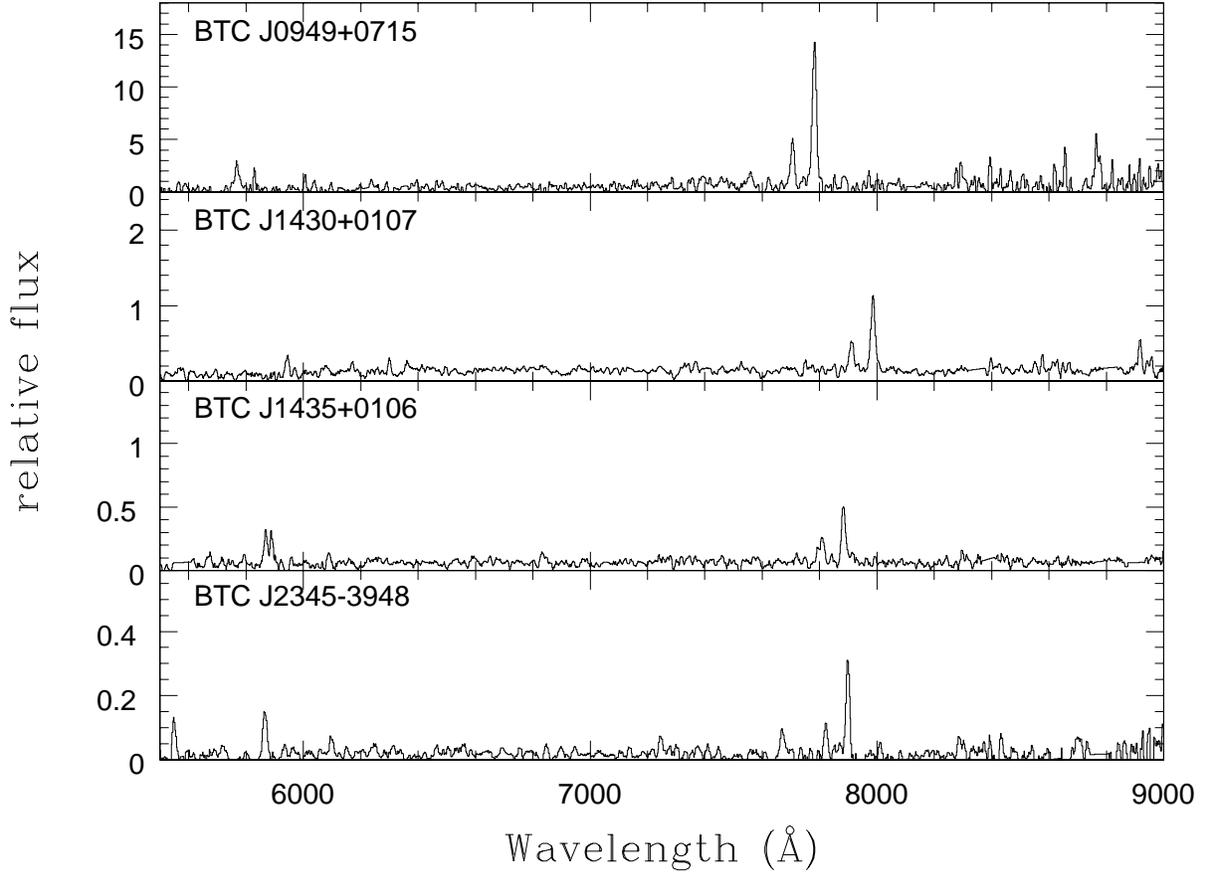}}
\figcaption[Monier.fig8.ps]{Spectra of four galaxies showing narrow emission lines due to 
[\ion{O}{3}] $\lambda\lambda$4959,5007 and [\ion{O}{2}] $\lambda$3727 at $z\approx0.6$.
Weak H$\beta$ $\lambda$4861 can also be seen in the top and bottom objects; 
it may be present in BTC40 J1435+0106 but unfortunately lies atop an atmospheric absorption band
where the noise is too large to allow its detection.
The upper three galaxies are compact sources identified as quasar candidates prior to spectroscopy.
The galaxy in the bottom panel was discovered serendipitously during spectroscopy of a 
nearby quasar candidate.  The spectra are smoothed by five pixels and are interpolated across regions
of poor sky subtraction.\label{cnelgs}}
\end{figure}

We report the discovery of three compact narrow emission line galaxies (CNELGs)
with 0.55 $< z <$ 0.6.  The positions, $I$ magnitudes and colors are provided for the CNELGs, 
in Table \ref{results}, finding charts are given in Figure \ref{cnelchart}, and the spectra are shown
in Figure \ref{cnelgs}.  The galaxies exhibit point-source profiles and 
[\ion{O}{3}] $\lambda\lambda$4959,5007 emission (Figure \ref{cnelgs}) shifted into the $I$ filter at 
$z \approx 0.6$, giving them colors similar to our quasar candidates.
All three galaxies have ($V-I$)$\approx$1.7, making them, along with the quasar
BTC40 1429+0119 ($V-I$=1.71), the bluest $(V-I)$ candidates in the sample
(see Figure \ref{colors}).
Contamination by such galaxies has been encountered by other surveys for quasars at
lower redshift (see, e.g. Hall \etal\ 1996 and Kennefick \etal\ 1997).
In addition to [\ion{O}{3}], the spectra show [\ion{O}{2}] $\lambda$3727 and
probable [\ion{Ne}{3}] $\lambda$3869, as well as H$\beta$ in one case.

BTC J0949+0715 is almost certainly a Seyfert 2 galaxy, based on the emission line 
diagnostics of Baldwin, Phillips, \& Terlevich (1981) and Rola, Terlevich, \& Terlevich (1997).  
Even though H$\beta$ is detected in this object, it is very weak relative
to [\ion{O}{3}] $\lambda$5007, and such a ratio is a good sign of AGN activity. 

The other galaxies appear to be starburst galaxies, though within 
the fairly large uncertainties, BTC J1430+0107 could be an AGN as well.

Finally, we note the serendipitous discovery of a galaxy at $z = 0.58$, as
measured from narrow \ion{O}{3}, \ion{O}{2}, and H$\beta$ emission lines seen
in the spectrum (Figure \ref{cnelgs}, bottom panel).  The galaxy is too faint for 
its morphology to be established from
the BTC image, but it lies 36\arcsec\ due north of quasar candidate BTC40 J2345-3948.  Since 
the slit was oriented N-S, spectra for the candidate and galaxy were
obtained simultaneously.

\section{Discussion}

In our high-redshift quasar survey we have found two $I<21.5$ quasars 
with $z = 4.6$ and $z = 4.8$ in 36 deg$^2$, proving the validity of the selection
technique.  Although our candidate selection is designed to be most sensitive for $z \gtrsim 4.8$, 
when the \lya\ emission line has moved into the $I$ filter, slightly lower-redshift ``lyman-break'' quasars 
may also enter the sample.  This was the case with the BTC2340$-$3949.

To compare our results with SDSS, Anderson \etal\ (2001) found 29 quasars 
with $z \gtrsim 4.5$ in $\approx$ 700 deg$^2$ of the SDSS commissioning data, for a surface density
of 1 quasar per 24 deg$^2$ to i$^*$ $\leq$ 20.5.  Four of these quasars had
$z > 5$, or 1 per 175 deg$^2$.  In the fall equatorial stripe of the
SDSS commissioning data, Fan et al. (2001a) found
5 quasars with  $4.5 \leq z \leq 4.77$  to $i^*$ $\lesssim$ 20 in 182 deg$^2$, or 1 in
36 deg$^2$.  Thus, our findings from the BTC40 survey are consistent
with the early SDSS results, although the BTC40 statistical base is
admittedly very small.  

More formally, the expected number of quasars predicted by a given
survey may be computed by numerically integrating the QLF determined by the
survey over the redshift and magnitude ranges of interest, and multiplying
by the efficiency of the survey.
Although we have not found enough quasars to derive a luminosity function,
our results can be compared to predictions based on the QLF
determined by the SDSS team. 

Fan \etal\ (2001b) used the 39 quasars from the SDSS commissioning data
to derive a QLF over the range $3.6 < z < 5$ and $-27.5 <$ \Mftf $< -25.5$, 
where \Mftf\ is the absolute continuum magnitude measured
at $\lambda = 1450$(1+$z$) \AA\ and calculated in the AB system.  Since our quasar spectra are
not spectrophotometric we scaled them to return the $I$ magnitudes previously measured
from the imaging data and then determined AB(1450(1+$z$)).  Assuming the continuum follows a power law 
with slope $\alpha$ = $-0.5$, then \Mftf = $-26.6$ for BTC40 J2340$-$3949 and \Mftf = $-26.8$
for BTC40 J1429+0119, and both quasars occupy the parameter space probed by SDSS.

When integrated over $4.5 < z < 5$, the luminosity function of Fan \etal\ (2001b) predicts 
a surface density of $\approx$0.026 quasars per square degree down to $i^{\prime} \approx 20.3$,
or $I=19.89$ using the conversion between the AB and conventional magnitude systems 
from Fukugita \etal\ (1996).  In the 36 deg$^2$ of the BTC40 survey, therefore, we
would expect to find $\approx$1 quasar with redshift $4.5 < z < 5$ and $I < 19.9$, and in fact
the two quasars we found fall into this category.
Furthermore, the absence of z $>$ 5 quasars in our sample to date is understandable, 
given their scarcity in the SDSS fields.  The SDSS luminosity function predicts
0.015 quasars per deg$^2$ with $5 < z < 6$ to $I\approx20$, or 
$<1$ quasars in the 36 deg$^2$ of our survey.  

The goal remains to determine the quasar luminosity function at fainter magnitudes. 
From the SDSS luminosity function over $4.5 < z < 5$ we would expect to find 10 quasars 
down to $I=21.5$ and 20 quasars to $I=22$ in our survey, while the QLF of SSG
predicts 15 quasars to $I=21.5$ and 35 to $I=22$.  Although we implemented candidate selection 
to $I=21.5$, in the end we attempted spectroscopy of only five candidates with $I > 21$.  None
of the resulting spectra were good enough to allow identification, but most likely the objects are stars given
the absence of any obvious emission lines.  Constraining 
the number of faint quasars in our survey may require additional $z$-band imaging data to reduce the scatter 
in the color-color diagram at faint magnitudes and enable candidate selection closer to the stellar locus.  

Follow-up spectroscopy on the $I > 21$ candidates will also necessitate the use of 8-10m  
telescopes to achieve a reasonable efficiency.  In addition to the selected candidates down to $I = 21.5$ 
remaining to be observed, we have identified 275 potential candidates with $21.5 < I < 22$.
Of course, effective use of time on the largest telescopes will require
additional work to keep cool stars from ending up in the candidate list.  As an example,
Fan \etal\ (2001c) used $J$-band photometry to pare L- and T-type dwarfs 
from their list of $i$-band dropouts in a search for quasars at $z \simeq 6$.

In summary, to date we have obtained spectroscopic observations of 
the brightest candidates in our survey for high-redshift quasars.
We have not yet found any quasars with $z > 5$, and in hindsight this
is not surprising given the results from SDSS.  But the SDSS results also predict our 
survey area will contain $\sim$5 quasars with $z > 5$ to $I = 21.5$, and 11 to $I = 22$.  
The next important step is to follow up on the 
fainter objects, which will require observations with 8-m class telescopes.

\section{Acknowledgements}

We are grateful to G. Bernstein, T. Tyson and the BTC team for their years of effort devoted to the
development and building of the BTC.  We would like to thank Alberto Conti, Rick Pogge 
and Darren Depoy for their assistance with the $z$-band imaging.  We also thank the CTIO and
AAO mountain staffs for their observing support.  This work was supported by NSF grant AST-9802658.

\clearpage


\clearpage

\input{Monier.tab1}

\input{Monier.tab2}

\input{Monier.tab3}

\input{Monier.tab4}

\end{document}

%% file: Monier.tab1.tex
\begin{deluxetable}{cccccc} 
\tablewidth{0in}

\tablenum{1}
\tablecaption{BTC40 Fields \label{fields}}
\tablehead{
 \colhead{}  &  \multicolumn{2}{c}{Field Center} & 
 \colhead{} & \colhead{}  &  \colhead{}  \cr
 \colhead{Field}   & \colhead{$\alpha$ (J2000)} & 
 \colhead{$\delta$ (J2000)} &  
 \colhead{$l^{\rm{II}}$} & \colhead{$b^{\rm{II}}$} & \colhead{Area\tablenotemark{a}
\ (deg$^2$)}  
 }
\startdata
F1 & 09 50 41 & \ 07 06 51\  & 229.4428 & \ 42.7567\ & 5.76 \\
F2 & 12 08 32 &  -19 44 16   & 289.2985 & \ 41.9954\ & 7.68 \\
F3 & 14 32 05 & \ 00 41 10\  & 349.5323 & \ 54.1881\ & 5.46 \\
F4 & 23 52 14 &  -40 12 00   & 342.3103 &  -72.0884  & 3.84 \\
F5 & 02 26 54 & \ 00 18 55\  & 166.7328 &  -54.2300  & 5.76 \\
F6 & 05 38 22 &  -28 12 44   & 232.3380 &  -27.4770  & 7.68 \\
\enddata
\tablenotetext{a}{Area covered by the {\it V}, {\it I}, and {\it z} filters.}
\end{deluxetable}

%% file: Monier.tab2.tex
\begin{deluxetable}{ccrccc} 
\tablewidth{0in}

\tablenum{2}
\tablecaption{Journal of CTIO Imaging Observations \label{journalofimaging}}
\tablehead{
 \colhead{}  &  \colhead{} & \colhead{} & \colhead{Exp Time\tablenotemark{a}}  &  
\colhead{5$\sigma$ Limiting} & \colhead{3$\sigma$ Limiting}  \\
 \colhead{Field}  &  \colhead{Filter} & \colhead{UT Date} & \colhead{(sec)}  
&  \colhead{Magnitude} & \colhead{Magnitude} 
}
\startdata
F1 & $V$ & 14,16-17 Mar 1997 & 300 & 23.6$\pm$0.2 & 24.2$\pm$0.2 \\
   & $I$ & 14,16-17 Mar 1997 & 150 & 22.2$\pm$0.2 & 22.7$\pm$0.2 \\
   & $z$ & 21-22 Feb 1999    & 300 & 22.2$\pm$0.2 & 22.7$\pm$0.2 \\
F2 & $V$ & 14-17 Mar 1997    & 300 & 24.3$\pm$0.1 & 24.8$\pm$0.1 \\
   & $I$ & 14-17 Mar 1997    & 150 & 22.7$\pm$0.1 & 23.2$\pm$0.1 \\
   & $z$ & 21-22 Feb 1999    & 300 & 22.6$\pm$0.1 & 23.1$\pm$0.1 \\
F3 & $V$ & 14-17 Mar 1997    & 300 & 24.1$\pm$0.2 & 24.6$\pm$0.2 \\
   & $I$ & 14-17 Mar 1997    & 150 & 22.4$\pm$0.1 & 23.0$\pm$0.1 \\
   & $z$ & 21-22 Feb 1999    & 300 & 22.5$\pm$0.2 & 23.0$\pm$0.2 \\
F4 & $V$ & 26 Nov 1997       & 300 & 24.0$\pm$0.1 & 24.6$\pm$0.1 \\
   & $I$ & 26-27 Nov 1997    & 150 & 22.2$\pm$0.1 & 22.8$\pm$0.1 \\
   & $z$ & 24-25 Nov 1998    & 300 & 22.3$\pm$0.2 & 22.9$\pm$0.2 \\
F5 & $V$ & 24,26-27 Nov 1997 & 300 & 23.9$\pm$0.2 & 24.4$\pm$0.2 \\
   & $I$ & 24,26-27 Nov 1997 & 150 & 22.2$\pm$0.2 & 22.8$\pm$0.2 \\
   & $z$ & 25-27 Nov 1998    & 300 & 22.2$\pm$0.1 & 22.8$\pm$0.1 \\
F6 & $V$ & 24-27 Nov 1997    & 300 & 24.1$\pm$0.1 & 24.7$\pm$0.1 \\
   & $I$ & 24,26-27 Nov 1997 & 150 & 22.4$\pm$0.2 & 22.9$\pm$0.2 \\
   & $z$ & 24-25 Nov 1998    & 300 & 22.4$\pm$0.2 & 22.9$\pm$0.2 \\

\enddata
\tablenotetext{a}{Exposure time per pointing within a field.}
\end{deluxetable}

%% file: Monier.tab3.tex
\begin{deluxetable}{rcrcc} 
\tablewidth{0in}

\tablenum{3}
\tablecaption{Journal of Spectroscopic Observations\label{journalofspec}}
\tablehead{
 \colhead{Telescope}  &  \colhead{Instrument} & \colhead{UT Date} 
 }
\startdata
CTIO 4-m & R-C Spec & 14-16 Nov 1999 \\
AAT  4-m & LDSS     & 2-6 Dec 1999   \\
AAT  4-m & LDSS     & 2-4 Apr 2000   \\
CTIO 4-m & R-C Spec & 9-11 May 2000  \\
AAT  4-m & LDSS     & 29-31 Aug 2000 \\
CTIO 4-m & R-C Spec & 16-17 Oct 2000 \\

\enddata
\end{deluxetable}

%% file: Monier.tab4.tex
\begin{deluxetable}{llcrcccccccc}
\rotate
\tablewidth{0in}

\tablenum{4}
\tablecaption{Results of QSO Candidate Spectroscopy\label{results}}
\tablehead{
 \colhead{No.} & \colhead{Name} &   \colhead{$\alpha$(2000)} &
 \colhead{$\delta$(2000)} & \colhead{{\it I} } 
& \colhead{\it V - I} & \colhead{\it I - z} & \colhead{Notes}
 }
\startdata
1	& BTC40 J0001$-$4025 &    00 01 20.6 & -40 25 16 &   21.18 & 4.27 &  0.09 \\
2	& BTC40 J0001$-$4023 &    00 01 34.3 & -40 23 21 &   20.74 & 2.33 &  0.09 \\
3	& BTC40 J0001$-$4025 &    00 01 50.5 & -40 24 25 &   21.04 & 2.48 & -0.02 \\
4	& BTC40 J0002$-$4019 &    00 02 14.8 & -40 19 52 &   20.78 & 1.97 &  0.02 \\
5	& BTC40 J0223+0028   &    02 23 07.7 &  00 28 09 &   20.46 & 2.86 &  0.21 \\
6       & BTC40 J0229+0023   &    02 29 58.7 &  00 23 05 &   20.32 & 2.18 &  0.11 \\
7       & BTC40 J0233$-$0008 &    02 33 42.2 & -00 08 20 &   20.48 & 2.05 &  0.14 \\
8       & BTC40 J0528$-$2711 &    05 28 47.4 & -27 11 30 &   19.80 & 1.90 & -0.13 \\
9 	& BTC40 J0529$-$2854 &    05 29 56.1 & -28 54 48 &   19.06 & 2.99 &  0.33 &  star  \\
10      & BTC40 J0537$-$2851 &    05 37 55.8 & -28 51 01 &   20.11 & 2.51 & -0.10 &  star  \\
11	& BTC40 J0544$-$2858 &    05 44 25.7 & -28 58 28 &   18.63 & 1.46 &  0.07 &  star  \\
12	& BTC40 J0948+0637   &    09 48 43.9 &  06 37 55 &   20.17 & 3.97 &  0.37 \\
13	& BTC40 J0949+0715   &    09 49 02.6 &  07 15 10 &   20.72 & 1.69 & -0.16 & ELG $z = 0.55$    \\
14	& BTC40 J0958+0733   &    09 58 59.5 &  07 33 57 &   20.04 & 2.01 &  0.19 &  star \\
15	& BTC40 J1202$-$2014 &    12 02 44.6 & -20 14 12 &   17.65 & 2.10 &  0.08 & M star \\
16	& BTC40 J1205$-$1918 &    12 05 02.9 & -19 18 09 &   21.47 & 3.89 &  0.15 \\
17	& BTC40 J1207$-$1916 &    12 07 47.7 & -19 16 39 &   20.65 & 2.02 &  0.04 \\
18	& BTC40 J1213$-$2022 &    12 13 14.8 & -20 22 28 &   19.60 & 2.42 &  0.20 & M star\\
19	& BTC40 J1216$-$1913 &    12 16 18.3 & -19 13 05 &   19.69 & 1.98 &  0.12 &  star \\
20	& BTC40 J1216$-$1934 &    12 16 27.2 & -19 34 48 &   20.73 & 2.30 &  0.03 \\
21	& BTC40 J1423+0046   &    14 23 28.5 &  00 46 02 &   21.00 & 2.26 &  0.06 \\
22	& BTC40 J1428+0119   &    14 28 49.8 &  01 19 33 &   20.77 & 2.34 &  0.05 \\
23	& BTC40 J1429+0119   &    14 29 26.5 &  01 19 54 &   19.35 & 1.71 & -0.08 & QSO $z = 4.84$ \\
24	& BTC40 J1429+0108   &    14 29 45.7 &  01 08 55 &   20.91 & 2.18 &  0.07 \\
25	& BTC40 J1429+0034   &    14 29 57.2 &  00 34 32 &   19.98 & 1.92 &  0.15 &  star \\
26	& BTC40 J1430+0028   &    14 30 09.7 &  00 28 11 &   20.50 & 3.60 &  0.30 \\
27	& BTC40 J1430+0107   &    14 30 10.9 &  01 07 22 &   19.88 & 1.68 &  0.08     & ELG $z = 0.59$ \\
28	& BTC40 J1434+0033   &    14 34 36.3 &  00 33 13 &   20.93 & 2.32 &  0.09 \\
29	& BTC40 J1435+0106   &    14 35 42.7 &  01 06 47 &   20.67 & 1.76 & -0.07 & ELG $z = 0.57$ \\
30	& BTC40 J1438+0043   &    14 38 24.8 &  00 43 23 &   20.61 & 3.35 &  0.29 \\
31	& BTC40 J2340$-$3949 &    23 40 25.7 & -39 49 32 &   19.40 & 2.05 &  0.12 & QSO $ z = 4.62$ \\
32	& BTC40 J2345$-$3948 &    23 45 27.5 & -39 48 42 &   21.38 & 3.92 &  0.09 & {}\tablenotemark{a} \\
33	& BTC40 J2348$-$4018 &    23 48 55.4 & -40 18 36 &   20.98 & 2.08 &  0.06 \\
34	& BTC40 J2350$-$4020 &    23 50 06.6 & -40 20 49 &   21.27 & 2.33 & -0.24 &  \\
35	& BTC40 J2354$-$4044 &    23 54 55.2 & -40 44 19 &   20.33 & 2.00 &  0.03 \\
36	& BTC40 J2356$-$3949 &    23 56 06.9 & -39 49 57 &   19.62 & 2.44 &  0.18 &  star \\
37      & BTC40 J2357$-$3944 &    23 57 10.1 & -39 44 15 &   21.23 & 2.27 & -0.12 \\
38	& BTC40 J2358$-$3951 &    23 58 23.2 & -39 51 04 &   19.60 & 2.12 &  0.17 &  star \\
39	& BTC40 J2358$-$3950 &    23 58 31.3 & -39 50 55 &   20.87 & 2.90 &  0.17 \\
40	& BTC40 J2358$-$4022 &    23 58 52.3 & -40 22 34 &   20.99 & 2.49 &  0.10 \\
\enddata   
\tablenotetext{a}{An emission-line galaxy at $z=0.58$ lies 36\arcsec\ north of the candidate.}
\end{deluxetable}

%% file: Monier.bbl
\begin{references}

\reference{}
Anderson, S.F., Fan, X., Richard, G.T., Schneider, D.P., Strauss, M.A, Van den Berk, D.E., 
Gunn, J.E. 2001, \aj, 122, 503

\reference{}
Baldwin, J.A., Phillips, M.M., \& Terlevich, R. 1981, \pasp, 93, 5

\reference{}
Becker, R.H., White, R.L., Gregg, M.D., Brotherton, M.S., Laurent-Muehleisen, S.A., \& Arav, N. 2000, \apj, 538, 72

\reference{}
Boyle, B.J., Shanks, T., \& Croom, S.M. 1995, \mnras, 276, 33

\reference{}
Fan, X., \etal\ 2000, \aj, 120, 1167

\reference{}
Fan, X., \etal\ 2001a, \aj, 121, 31

\reference{}
Fan, X., \etal\ 2001b, \aj, 121, 54

\reference{}
Fan, X., \etal\ 2001c, \aj, 122, 2833

\reference{}
Fukugita, M., Ikchikawa, T., Gunn, J.E., Doi, M., Shimasaku, K., \& Schneider, D.P. 1996, \aj, 111, 1748

\reference{}
Glazebrook, K., \& Bland-Hawthorn, J. 2001, \pasp, 780, 197

\reference{}
Gunn, J. E., \& Stryker, L. L. 1983, \apjs, 52, 121

\reference{}
Gunn, J.E., \etal\ 1998, \aj, 116, 3040

\reference{}
Haehnelt, M.G., Natarajan, P., \& Rees, M.J. 1998, \mnras, 300, 817

\reference{}
Haehnelt, M.G., \& Kauffmann, G. 2000, \mnras, 318, L35

\reference{}
Hall, P.B., Osmer, P.S., Green, R.F., Porter, A.C., \& Warren, S.J. 1996, \aj, 462, 614

\reference{}
Jarvis, J.F. \& Tyson, J.A. 1981, \aj, 86, 426

\reference{}
Kellerman, K.I., Sramek, R., Schmidt, M., Shaffer, D.B., \& Green, R. 1989, \aj, 98, 1195 

\reference{}
Kennefick, J.D., Djorgovski, S.G., \& de Carvalho 1995, \aj, 110, 2553

\reference{}
Kennefick, J.D., Djorgovski, S.G., \& Meylan, G. 1996, \aj, 111, 1816

\reference{}
Kennefick, J.D., Osmer, P.S., Hall, P.B., \& Green, R.F. 1997, \aj, 114, 2269

\reference{}
Madau, P., Haardt, F., \& Rees, M.J. 1999, \apj, 514, 648

\reference{}
Monet, D., Bird A., Canzian, B., Dahn, C., Guetter, H., Harris, H., Henden, A., Levine, S., 
Luginbuhl, C., Monet, A. K. B., Rhodes, A., Riepe, B., Sell, S., Stone, R., Vrba, F., \& 
Walker, R. 1998, The USNO-A2.0 Catalogue, (U.S. Naval Observatory, Washington DC). 

\reference{}
Prandoni, I., Gregorini, L., Parma, P., de Ruiter, H.R., Vettolani, G., Wieringa, W.H., 
\& Ekers, R.D. 2000, \aaps, 146, 41

\reference{} 
Schmidt, M., Schneider, D. P., \& Gunn, J. E. 1995, \aj, 110, 68 (SSG)

\reference{}
Sharp, R.G., McMahon, R.G., Irwin, M.J., \& Hodgkin, S.T. 2001, \mnras, 326, L45

\reference{}
Strecker, D. W., Erickson, E. F., \& Whittenborn, F. C. 1979, \apjs, 41, 501

\reference{}
Rola, C.S., Terlevich, E., \& Terlevich, R.J. 1997, \mnras, 289, 419

\reference{}
Tyson, Bernstein, Blouke, \& Lee 1992, SPIE 1656, 400

\reference{}
Valdes, F. 1982, {\it SPIE Proc. on Instrumentation in Astronomy IV}, 331, 465

\reference{} 
Warren, S. J., Hewett, P. C., \& Osmer, P. S. 1994, \apj, 421, 412 (WHO)

\reference{}
Weir, N., Fayyad, U.M., Djorgovski, S.G., \& Roden, J. 1995, \pasp, 107, 1243

\reference{}
Wittman, D.M., Tyson, J.A., Bernstein, G.M., Lee, R.W., Dell'Antonio, I.P., Fischer, P., Smith, D.R., \&
Blouke, M.M. 1998, SPIE, 3355, 626



\end{references}
